\documentclass[preprint,nofootinbib, superscriptaddress, amsmath,amssymb,aps, prd]{revtex4-1}

\usepackage{graphicx}
\usepackage{dcolumn}
\usepackage{bm}

\usepackage{amsmath}
\usepackage{bm}
\usepackage{slashed}
\usepackage{epsfig}
\usepackage{amsfonts}
\usepackage{epstopdf}
\usepackage{color}
\usepackage{float}

\begin{document}

\preprint{}

\title{Relativistic corrections to prompt double charmonium hadroproduction near threshold}

\author{Zhi-Guo He}
\affiliation{Department of Physics and Electronics, School of Mathematics and Physics, Beijing University of Chemical Technology, Beijing 100029, China}
\affiliation{{II.} Institut f\"ur Theoretische Physik, Universit\"at Hamburg,
Luruper Chaussee 149, 22761 Hamburg, Germany}
\affiliation{Institut f\"ur Theoretische Physik, 
  Universit\"at Regensburg,
  Universit\"atsstra{\ss}e~31,
  93053 Regensburg, Germany}
\author{Xiao-Bo Jin}
\affiliation{Center of Advanced Quantum Studies, Department of Physics, Beijing Normal University, Beijing 100875, China}
\author{Bernd A. Kniehl}
\affiliation{{II.} Institut f\"ur Theoretische Physik, Universit\"at Hamburg,
Luruper Chaussee 149, 22761 Hamburg, Germany}

\date{\today}

\begin{abstract}
  We calculate the relativistic corrections to prompt $J/\psi$ pair and $J/\psi+\psi(2S)$ hadroproduction through the color-singlet channel within the framework of nonrelativistic QCD (NRQCD) factorization. The short-distance coefficients are obtained by matching full-QCD and NRQCD calculations at the partonic level, in which both squared amplitude and phase space are expanded in $v^2$. We find that such an expansion of the phase space spoils the convergence of NRQCD factorization near the production threshold. To fix this problem, we propose to modify the matching between full QCD and NRQCD by adopting the physical phase space. In this modified approach, the theoretical uncertainties due to the choice of charm quark mass $m_c$ are largely reduced and the overall agreement of our predictions with LHC data is significantly improved, both as for total and differential cross sections.
\end{abstract}

\maketitle


\section{Introduction}

The production mechanism of heavy quarkonium, the bound state of a heavy quark-antiquark pair, has been a long-standing puzzle in high-energy particle physics ever since the discovery of the first charmonium state $J/\psi$ about half a century ago.
Due to the hierarchy of energy scales in heavy-quarkonium production, the nonrelativistic QCD (NRQCD) factorization approach \cite{Bodwin:1994jh}, which was established on top of the NRQCD effective field theory \cite{Caswell:1985ui}, has been generally viewed as the most suitable tool to study this process.
In NRQCD factorization, the cross sections of heavy-quarkonium production are factorized into products of short-distance coefficients (SDCs) and long-distance matrix elements (LDMEs).
The SDCs describe the creation of the heavy-quark pair and can be calculated perturbatively in powers of the strong-coupling constant $\alpha_s$.
The supposedly universal LDMEs represent the non-perturbative evolution of heavy-quark pairs into the heavy-quarkonium states and scale as powers of $v$ according to velocity scaling rules \cite{Lepage:1992tx}, where $v$ is the relative quark velocity in the center-of-mass frame.
In this way, the theoretical predictions are organized as double expansions in $\alpha_s$ and $v$.
A key feature of NRQCD factorization is the incorporation of all possible Fock states, including both color singlet (CS) and color octet (CO) ones.
The possible dominance of CO states, dubbed CO mechanism, greatly improves our understanding of the heavy-quarkonium production.
However, the verification of LDME universality in global fits at next-to-leading order (NLO) to experimental data still poses a serious challenge, especially in the most important case of the $J/\psi$ meson (see Ref.~\cite{Lansberg:2019adr} for a recent review). 

In recent years, $J/\psi$ associated hadroproduction with another quarkonium state, a $W$, or a $Z$ boson \cite{Butenschoen:2022wld} have attracted considerable attention.
Among them, prompt $J/\psi$ pair production is of particular interest not only because of doubled LDME sensitivity, but also because this provides a laboratory for the study of double parton scattering (DPS), which is important at large invariant mass $m_{\psi\psi}$ and large rapidity separation $|\Delta y|$, and the extraction of the key DPS parameter $\sigma_{\mathrm{eff}}$ \cite{Kom:2011bd}.
Moreover, this is the background for the newly discovered tetra-quark state $X(6900)$ \cite{LHCb:2020bwg,ATLAS:2023bft,CMS:2023owd}, which mainly decays into $J/\psi$ pairs.

On the experimental side, prompt double $J/\psi$ hadroproduction has been measured by the D0 Collaboration \cite{D0:2014vql} at the Fermilab Tevatron and by the LHCb~\cite{LHCb:2011kri,LHCb:2016wuo}, CMS~\cite{CMS:2014cmt}, and ATLAS Collaborations~\cite{ATLAS:2016ydt} at the CERN LHC.
Under the assumption that the single parton scattering (SPS) contribution can be neglected in the large-$|\Delta y|$ region, both the D0 and LHCb Collaborations were able to extract $\sigma_{\mathrm{eff}}$, yet with different results.
Recently, the LHCb Collaboration meticulously measured total and differential cross sections of inclusive $2J/\psi$ \cite{LHCb:2023ybt} and $J/\psi+\psi(2S)$ \cite{LHCb:2023wsl} hadroproduction at center-of-mass energy $\sqrt{s}=13$~TeV, separately for SPS and DPS in the $2J/\psi$ case, enabling meaningful comparisons with NRQCD predictions.

On the theoretical side, the idea to probe LDMEs in double $J/\psi$ hadroproduction was first raised by Barger {\it et al.}\ in Ref.~\cite{Barger:1995vx}, where the fragmentation contribution to the $2c\bar{c}({}^3S_1^{[8]})$ production channel was considered. Later on, it was found that the leading-order (LO) CS channel $gg\to2c\bar{c}({}^3S_1^{[1]})$ mainly contributes at small and moderate values of transverse momentum $p_T$ \cite{Qiao:2002rh,Li:2009ug,Ko:2010xy}, where it is largely enhanced at NLO \cite{Sun:2014gca}, while in the large-$m_{\psi\psi}$ and large-$|\Delta y|$ regions, contributions involving $t$-channel gluon exchange dominate by orders of magnitude \cite{He:2015qya}.
Upon including all possible LO contributions, the SPS predictions in the large-$m_{\psi\psi}$ and large-$|\Delta y|$ regions were found to underestimate experimental measurements by more than one order of magnitude \cite{He:2015qya}.
To reduce the large gap, higher-order QCD corrections to some channels \cite{He:2019qqr,Lansberg:2019fgm,Sun:2023exa} were calculated, and large logarithm of the form $(\alpha_s\ln|s/t|)^n$ were resummed using high-energy factorization~\cite{He:2019qqr}.
Notice that, in the case of double $P$-wave Fock state production, a new type of infrared divergences causes NRQCD factorization as we know it to break down \cite{He:2018hwb} pushing complete NLO NRQCD predictions out of reach for the time being.
Besides experimental determinations \cite{D0:2014vql,LHCb:2011kri,LHCb:2016wuo},
also theoretical groups, armed with advanced knowledge of SPS contributions, 
fitted $\sigma_\mathrm{eff}$ to experimental data finding different results \cite{Lansberg:2014swa,Prokhorov:2020owf}.
For a recent review of theoretical progress on prompt double $J/\psi$ hadroproduction, we refer to Ref.~\cite{He:2021oyy}.

While NRQCD predictions for prompt double $J/\psi$ hadroproduction were generally found to undershoot the experimental data in the large-$m_{\psi\psi}$ region, the situation is very different close to the production threshold, $m_{\psi\psi}\agt2m_{J/\psi}$ \cite{He:2015qya}.
In fact, the LHCb measurement at center-of-mass energy $\sqrt{s}=7~\mathrm{TeV}$ undershoots the NRQCD prediction by almost one order of magnitude in the first bin, $6~\mathrm{GeV}<m_{\psi\psi}<7~\mathrm{GeV}$ \cite{He:2015qya}.
Close to threshold, the CO contribution is negligibly small and the QCD-corrected CS channel $gg\to2c\bar{c}({}^3S_1^{[1]})$ significantly falls short of the LHCb data. 

Besides NLO QCD corrections, higher-order relativistic corrections may also have an important impact given that $v^2\approx\alpha_s(2m_c)$ for charmonium.
This is supported, {\it e.g.}, by studies of $J/\psi$ associated production with $\eta_c$ mesons, charmed \cite{He:2007te} and light hadrons \cite{He:2009uf,Jia:2009np} in $e^+e^-$ annihilation and by $J/\psi$ yield \cite{Xu:2012am,He:2014sga} and polarization \cite{He:2015gla} in single inclusive photo- and hadroproduction.

Higher-order relativistic correction to inclusive double $J/\psi$ hadroproduction were first studied in the context of the relativistic quark model and were found to reduce the LO total cross section by a factor of 6, down to a value undershooting the LHCb measurement \cite{LHCb:2011kri} by more than a factor of 3 \cite{Martynenko:2012tf}.
In the analogous study for differential cross sections in the NRQCD factorization framework \cite{Li:2013csa}, the relativistic corrections were found to be just a few percent, and it was argued that their effect on the phase space is suppressed at the hadron level by an additional factor of $1/\ln(\sqrt{s}/m_T)$, where $m_T=\sqrt{m_{J/\psi}^2+p_T^2}$ is the $J/\psi$ transverse mass. 

This ambiguous situation is unsatisfactory and together with the new high-precision data from LHCb \cite{LHCb:2023ybt,LHCb:2023wsl} motivates us to revisit Ref.~\cite{Li:2013csa}, which has not been confirmed yet.
In the following, we thus perform an independent calculation of the NLO relativisitic corrections to inclusive double charmonium hadroproduction via SPS in the 
NRQCD factorization framework, placing special emphasis on the threshold region, which actually requires a careful treatment of the phase space.

The remainder of this paper is organized as follows.
In Sec.~\ref{sec:frame}, we describe the calculation of SCDs in detail.
In Sec.~\ref{sec:ph}, we present our numerical results for the NRQCD predictions including relativistic corrections, both for $2J/\psi$ and $J/\psi+\psi(2S)$ production, and compare them with respective LHCb data \cite{LHCb:2023ybt,LHCb:2023wsl}.
Our conclusions are summarized in Sec.~\ref{sec:summary}.

\section{Analytic Results}\label{sec:frame}

According to the factorization theorem of the collinear parton model, the hadroproduction rate of inclusive double prompt $J/\psi$ is given by
\begin{equation}
  \sigma(A+B\to 2J/\psi+X) =\sum_{i,j}
  \int\mathrm{d}x_1\mathrm{d}x_2
  f_{i/A}(x_1)f_{j/B}(x_2) \hat{\sigma}(i+j\to 2J/\psi+X)\,,
\end{equation}
where $f_{i/A}(x)$ is the parton distribution function (PDF) of parton $i$ in hadron $A$. In NRQCD factorization through relative order $\mathcal{O}(v^2)$, the partonic cross section can be further factorized as 
\begin{eqnarray}
\lefteqn{\hat{\sigma}(i+j\to 2J/\psi+X)
 =\sum_{m,n,H_1,H_2} 
\left(\frac{F^{ij}(m,n)}{m_c^{d_{\mathcal{O}(m)}-4}m_c^{d_{\mathcal{O}(n)}-4}}
\langle \mathcal{O}^{H_1}(m)\rangle \langle \mathcal{O}^{H_2}(n)\rangle
\vphantom{\frac{G_{1}^{ij}(m,n)}{m_c^{d_{\mathcal{P}(m)}-4}m_c^{d_{\mathcal{O}(n)}-4}}}
\right.}\nonumber\\
&&{}+\left.\frac{G_{1}^{ij}(m,n)}{m_c^{d_{\mathcal{P}(m)}-4}m_c^{d_{\mathcal{O}(n)}-4}}
  \langle \mathcal{P}^{H_1}(m)\rangle \langle \mathcal{O}^{H_2}(n)\rangle
  +\frac{G_{2}^{ij}(m,n)}{m_c^{d_{\mathcal{O}(m)}-4}m_c^{d_{\mathcal{P}(n)}-4}}
  \langle \mathcal{O}^{H_1}(m)\rangle \langle \mathcal{P}^{H_2}(n)\rangle
  \right)\nonumber\\
&&{}\times
  \mathrm{Br}(H_1\to J/\psi+X)\mathrm{Br}(H_2\to J/\psi+X)\,,
\end{eqnarray}
where $\mathcal{O}^{H}(m)$ and $\mathcal{P}^{H}(m)$ are the LO and $\mathcal{O}(v^2)$ four-quark operators pertaining to the transition $n\to H$ with dimension $d_{\mathcal{O}(m)}$ and $d_{\mathcal{P}(m)}$, respectively; $F^{ij}(m,n)$, $G_{1}^{ij}(m,n)$, and $G_{2}^{ij}(m,n)$ are the corresponding SDCs of the partonic subprocess $i+j\to c\bar{c}(m)+c\bar{c}(n)+X$, and $\mathrm{Br}(H \rightarrow J/\psi+X)$ is the branching fraction, which equals unity for $H=J/\psi$.
Near threshold, the $J/\psi$ pair is mainly produced through the CS channel of the gluon fusion process, $gg\to 2c\bar{c}(^3S_1^{[1]})$, on which we concentrate in the following.
This also includes the feed-down contribution from $\psi(2S)$.

The relevant four-quark operators for the production of $H=J/\psi,\psi(2S)$ are
defined as~\cite{Bodwin:1994jh}
\begin{eqnarray}
  \mathcal{O}^{H}({}^3{S}_1^{[1]})&=&\chi^{\dagger}\bm{\sigma}^i\psi(a_H^\dagger 
  a_H)\psi^{\dagger}\bm{\sigma}^i\chi\,,\nonumber\\
  \mathcal{P}^{H}({}^3{S}_1^{[1]})&=&\chi^{\dagger}\bm{\sigma}^i\psi(a_H^\dagger 
  a_H)\psi^{\dagger}\bm{\sigma}^i\left(-\frac{i}
  {2}\overleftrightarrow{\bm{\mathrm{D}}}\right)^2\chi+\mathrm{h.c.}\,,
\end{eqnarray}
where $\psi$ ($\chi$) is the Pauli spinor that annihilates (creates) a heavy quark (antiquark), 
$\bm{\sigma}^i$ ($i=1,2,3$) are the Pauli matrices, and $\overleftrightarrow{\bm{\mathrm{D}}}=
\overleftarrow{\bm{\mathrm{D}}}-\overrightarrow{\bm{\mathrm{D}}}$, with 
$\overrightarrow{\bm{\mathrm{D}}}$ being the space component of the covariant derivative $D^{\mu}$.

In NRQCD factorization, the SDCs are the same for the $J/\psi$ and $\psi(2S)$ mesons, since the Fock states carry the same quantum numbers.
Therefore, we only need to calculate  
$F({^3S_1^{[1]},^3S_1^{[1]}})$ and $G_{1,2}({^3S_1^{[1]},^3S_1^{[1]}})$ perturbatively, through the matching condition between full QCD and the NRQCD,
\begin{eqnarray}
&&\left.\hat{\sigma}(g+g\to  (c\bar{c})_1+(c\bar{c})_2)\right|_\mathrm{pert\, QCD}
  =\left[\frac{F({}^3S_1^{[1]},{}^3S_1^{[1]})}{m_c^4}\langle 0|\mathcal{O}^{(c\bar{c})_1}({}^3S_1^{[1]})|0\rangle
  \langle 0|\mathcal{O}^{(c\bar{c})_2}({}^3S_1^{[1]})|0\rangle\right.\nonumber\\
  &&{}+
  \frac{G_{1}({}^3S_1^{[1]},{}^3S_1^{[1]})}{m_c^6}
  \langle 0|\mathcal{P}^{(c\bar{c})_1}({}^3S_1^{[1]}) |0\rangle
  \langle 0|\mathcal{O}^{(c\bar{c})_2}({}^3S_1^{[1]}) |0\rangle\nonumber\\
  &&{}+\left. \frac{G_{2}({}^3S_1^{[1]},{}^3S_1^{[1]})}{m_c^6} 
  \langle 0|\mathcal{O}^{(c\bar{c})_1}({}^3S_1^{[1]}) |0\rangle
  \langle 0|\mathcal{P}^{(c\bar{c})_2}({}^3S_1^{[1]}) |0\rangle\right]_\mathrm{pert\, NRQCD}\,.
 \label{eq:match} 
\end{eqnarray}
The left-hand side of Eq.~\eqref{eq:match} can be computed easily by implementing the covariant-projection method \cite{Bodwin:2002cfe}.
In this method, the product of Dirac spinors $v(p_{\bar{c}})\bar{u}(p_c)$ is written in a Lorentz-invariant form to all orders of $v^2$ after projection onto spin singlet, $S=0$, or spin triplet, $S=1$, configurations.
The four-momenta of the $c$ and $\bar{c}$ quarks of the $(c\bar{c})$ pair in an arbitrary reference frame can be related to those in the $(c\bar{c})$ rest frame via an appropriate Lorentz boost matrix $L^\mu_{\phantom{\mu}\nu}$ as 
\begin{eqnarray}
  p_{c}&=&
  \frac{P}{2}+q=L
  \left(\frac{P_r}{2}+\bm{q}\right)\,,\nonumber
  \\
  p_{\bar{c}}&=&
  \frac{P}{2}-q=L
  \left(\frac{P_r}{2}-\bm{q}\right)\,,
\end{eqnarray}
where $P_r^\mu=(2E_q,\bm{0})$, $E_q=\sqrt{\bm{q}^2+m_c^2}$, and $2\bm{q}$ is the relative three-momentum of the $c$ and $\bar{c}$ quarks in the $(c\bar{c})$ rest frame.
Then, in an arbitrary frame, the projection on spin triplet reads~\cite{Bodwin:2002cfe}
\begin{equation}
  \sum_{\lambda_1,\bar{\lambda}_1} v(p_{\bar{c}},\bar{\lambda}_1)\bar{u}(p_c,\lambda_1)\left\langle
  \frac{1}{2},\lambda_1;\frac{1}{2},\bar{\lambda}_1|1,S_{z}\right\rangle  =
  \frac{1}{\sqrt{2}(E_q+m_c)}  \left( \slashed{p}_{\bar{c}}-m_c \right)
  \slashed{\epsilon} \frac{\slashed{P}+2E_q}{4E_q} \left(\slashed{p}_c+m_c \right)\,,
\end{equation}
where $\epsilon^\mu$ is the polarization vector of the spin triplet state and the Dirac spinors are normalized as $\bar{u}u=-\bar{v}v=2m_c$. 
In this way, the full QCD scattering amplitude can be expressed as  
\begin{equation}
  \mathcal{M}\left(g+g\to 
  (c\bar{c})_1(^3S_1^{[1]})+(c\bar{c})_2(^3S_1^{[1]})\right)=
  \sqrt{\frac{m_c}{E_{q_1}}}\sqrt{\frac{m_c}{E_{q_2}}} A(q_1,q_2)\,,
\end{equation}
where the factor $\sqrt{m_c/E_{q_i}}$ compensates the relativistic normalization of the $(c\bar{c})_i$ pair and 
\begin{eqnarray}
A(q_1,q_2)&=&
  \sum_{\lambda_i\bar{\lambda}_i k_il_i} 
  \left\langle \frac{1}{2},\lambda_1;\frac{1}{2},\bar{\lambda}_1 |1,S_{1z}
  \right\rangle
  \left\langle 3,k_1;\bar{3},l_1|1 \right\rangle 
  \left\langle \frac{1}{2},\lambda_2;\frac{1}{2},\bar{\lambda}_2 |1,S_{2z}
  \right\rangle
  \left\langle 3,k_2;\bar{3},l_2 |1\right\rangle 
  \nonumber\\
  &&{}\times\mathcal{A}(g+g\to c_{\lambda_1,k_1}(p_{c_1})+\bar{c}_{\bar{\lambda}_1,l_1}(p_{\bar{c}_1})+c_{\lambda_2,k_2}(p_{c_2})+\bar{c}_{\bar{\lambda}_2,l_2}(p_{\bar{c}_2}))\,,
\label{eq:a}
\end{eqnarray}
with $\mathcal{A} \big{(}g+g\to c_{\lambda_1,k_1}(p_{c_1})+\bar{c}_{\bar{\lambda}_1,l_1}
(p_{\bar{c}_1})+c_{\lambda_2,k_2}(p_{c_2})+\bar{c}_{\bar{\lambda}_2,l_2}(p_{c_2})\big{)} $ being the standard Feynman amplitude and $\langle 3,k_i;\bar{3},l_i |1 \rangle=\delta_{k_il_i}/\sqrt{3}$ being the Clebsch-Gordan coefficient in SU(3) color space for the $c\bar{c}$ pair to be projected onto the CS state. 

To expand Eq.~(\ref{eq:a}) as a double series in $q_1$ and $q_2$, it is convenient to define
\begin{equation}
A_{\alpha_1\cdots\alpha_m,\beta_1\cdots\beta_n}(0,0)\equiv\frac{\partial^{m+n}A(q_1,q_2)}{\partial q_1^{\alpha_1}\cdots\partial q_1^{\alpha_m}\partial q_2^{\beta_1}\cdots\partial q_2^{\beta_n}}\bigg{|}_{q_1=q_2=0}\,,
\end{equation}
so that
\begin{eqnarray}
  A(q_1,q_2)&=&A(0,0)+q_1^{\alpha_1}A_{\alpha_1,0}+q_2^{\beta_1}A_{0,\beta_1}(0,0)
  \nonumber\\
&&{}+\frac{1}{2}q_1^{\alpha_1}q_1^{\alpha_2}A_{\alpha_1\alpha_2,0}(0,0)+\frac{1}{2}q_2^{\beta_2}q_2^{\beta_2}A_{0,\beta_1\beta_2}(0,0)+\cdots\,.
\end{eqnarray}
For $S$-wave production, only even powers of $q_i$ contribute.
To extract the higher-order corrections in $v^2$, we need to further decompose the higher-rank tensor products into the irreducible representations and retain the $S$-wave terms.
At $\mathcal{O}(v^2)$, we have
\begin{equation}
 q^{\mu} q^{\nu}= \frac{|\bm{q}|^2}{3} \left(-g^{\mu \nu}+ \frac{P^{\mu} P^{\nu}}{4E_q^2}\right)\equiv\frac{|\bm{q}|^2}{3} \Pi^{\mu\nu}\,.
\end{equation}
Through $\mathcal{O}(v^2)$, we thus have
\begin{eqnarray}
\mathcal{M}\left(g+g\to 
c\bar{c}_1(^3S_1^{[1]})+c\bar{c}_2(^3S_1^{[1]})\right)&=&
\sqrt{\frac{m_c}{E_{q_1}}}\sqrt{\frac{m_c}{E_{q_2}}}
\left[ A(0,0) +\frac{|\bm{q}_1|^2}{6} \Pi_1^{\alpha_1\alpha_2}A_{\alpha_1\alpha_2,0}(0,0)
  \right.\nonumber\\
 &&{}+\left.
  \frac{|\bm{q}_2|^2}{6} \Pi_2^{\beta_1\beta_2}A_{0,\beta_1\beta_2}(0,0)\right]\,.
\label{eq:amplitude}
\end{eqnarray}

In Eq.~(\ref{eq:amplitude}), the second and third terms within the square brackets are already of $\mathcal{O}(v^2)$. 
However, the first term $A(0,0)$ still depends on $\bm{q}_1^2$ and $\bm{q}_2^2$ implicitly via $P_i^2=4E_{q_i}^2\equiv m_{H_i}^2$ ($i=1,2$) and the Mandelstam variables 
\begin{equation}
  t=(k_1-P_1)^2=(k_2-P_2)^2\,,\qquad  
  u=(k_1-P_2)^2=(k_2-P_1)^2\,,
\end{equation}
where $k_1$ and $k_2$ are the four-momenta of the incoming gluons.
Notice that $s=(k_1+k_2)^2$ is independent of $\bm{q}_1^2$ and $\bm{q}_2^2$.
To expand $t$ and $u$ in $v^2$, we find it is more convenient to introduce two new variables, 
\begin{equation}\label{eq:tu}
  \hat{t}\equiv t+\frac{s-m_{H_1}^2-m_{H_2}^2}{2}\,,\qquad
  \hat{u}\equiv u+\frac{s-m_{H_1}^2-m_{H_2}^2}{2}\,.
\end{equation}
The two-body phase space measure thus takes a very simple form, 
\begin{equation}\label{eq:ps2}
  \mathrm{d}\Phi_2=
  \frac{\mathrm{d}\hat{t}
  \mathrm{d}\hat{u}}{8\pi s}
  \,\delta(\hat{t}+\hat{u})\,.
\end{equation}
In the nonrelativistic limit, $\bm{q}_{1,2}^2\to 0$, we have $t\to t_0$ and
$u\to u_0$ with $ s + t_0 + u_0=8m_c^2$.
Analogously to Eq.~\eqref{eq:tu}, we introduce 
\begin{equation}\label{eq:t0u0}
  \hat{t}_0 \equiv t_0+\frac{s-8m_c^2}{2}\,,\qquad
  \hat{u}_0 \equiv u_0+\frac{s-8m_c^2}{2}\,.
\end{equation}
It is straightforward to see that
\begin{equation}\label{eq:kfactor}
\frac{\hat{t}}{\hat{t}_0}  = \frac{\hat{u}}{\hat{u}_0}  =
 \sqrt{ \frac{\lambda (s,m_{H_1}^2,m_{H_2}^2)}{\lambda (s,4m_c^2,4m_c^2)}} \equiv k\,,
\end{equation}
where
\begin{equation}
  \lambda(a,b,c) = a^2+b^2+c^2-2ab-2bc-2ac\,.
\end{equation}
Thus, the two-body phase space integral in Eq.~(\ref{eq:ps2}) can also be re-expressed in $\hat{t}_0$ and $\hat{u}_0$ as
\begin{equation}\label{eq:ps20}
\mathrm{d}\Phi_2 = k \frac{\mathrm{d}\hat{t}_0 \mathrm{d}\hat{u}_0} {8\pi s}
  \delta(\hat{t}_0+\hat{u}_0)=k \mathrm{d}\Phi_{20}\,.
\end{equation}
With the help of Eqs.~(\ref{eq:tu}), (\ref{eq:t0u0}), and (\ref{eq:kfactor}), both $A(0,0)$ and the two-body phase space measure in Eq.~(\ref{eq:ps20}) can be easily expanded in $v^2$, too. 

We are now in a position to present the master formulas for the SDCs in the matching condition in Eq.~(\ref{eq:match}),
\begin{eqnarray}
\frac{F({}^3S_1^{[1]},{}^3S_1^{[1]})}{m_c^4} &=&\frac{1}{2s}\int d\Phi_{20} |M|^2\,,
\nonumber\\
\frac{G_1({}^3S_1^{[1]},{}^3S_1^{[1]})}{m_c^6} &=&\frac{G_2({}^3S_1^{[1]},{}^3S_1^{[1]})}{m_c^6} =
\frac{1}{2s}\int d\Phi_{20} (K|M|^2+|N|^2)\,,
\end{eqnarray}
with 
\begin{eqnarray}
|M|^2&=&\left.\overline{\sum}|A(0,0)|^2\right|_{\bm{q}_1^2=\bm{q}_2^2=0}\,,\nonumber\\
|N|^2&=&\left[\frac{\partial}{\partial \bm{q}_1^2 }\left (\frac{m_c}{E_{q_1}}\overline{\sum}|A(0,0)|^2\right)+
\frac{1}{3}\overline{\sum}|\Pi_{1}^{\alpha_1\alpha_2}
\mathrm{Re}\left[A^{\ast}(0,0)A_{\alpha_1\alpha_2,0}\right]\right]_{\bm{q}_1^2=\bm{q}_2^2=0}\,,
\end{eqnarray}
where the symbol $\overline{\sum}$ implies summing over all polarizations in the initial and final states and then averaging the spins and colors of the incoming partons and the outgoing $c\bar{c}$ Fock states.
The factor $K=-4/(s-16m_c^2)$ originates from the expansion of the $k$ factor in Eq.~(\ref{eq:ps20}) and is divergent at threshold, although the complete phase space integral is still convergent.  

From the explicit expression of the $k$ factor,
\begin{equation}
k=\sqrt{1-\frac{8\bm{q}_1^2+8\bm{q}_2^2}{s-16m_c^2}+\frac{16(\bm{q}_1^2-\bm{q}_2^2)^2}{s(s-16m_c^2)}}\,,
\end{equation}
we observe that the actual expansion parameter for the phase space is not $v^2=\bm{q}^2/m_c^2$, but rather $8m_c^2v^2/(s-16m_c^2)$, which is not small near threshold.
We hence conclude that, for $J/\psi$ pair production near threshold, the usual fixed-order NRQCD predictions will not be reliable unless the numerical value of 
$8m_c^2/(s-16m_c^2)$ is of order unity or smaller.
To overcome this problem, we alternatively choose the physical phase space in 
our numerical evaluation, without the expansion in $v^2$, and replace $t_0$ and $u_0$ by their physical counterparts, $t_p$ and $u_p$, using a similar relation as Eq.~(\ref{eq:kfactor}).
The numerical consequence of this treatment is addressed in Sec.~\ref{sec:ph}.

We use the program packages Feyn~Arts \cite{Kublbeck:1990xc} to generate Feynman diagrams and Feyn~Calc \cite{Mertig:1990an} to handle Dirac algebra, color indices, and expansion of the squared amplitude.
When we drop the relativistic corrections to the phase space, our result for $G_1({}^3S_1^{[1]},{}^3S_1^{[1]})$ agrees with that in Ref.~\cite{Li:2013csa} except for a factor of 1/2 due to the identical-boson symmetry.
We employ the program package CUBA \cite{Hahn:2004fe} to perform the numerical phase space integrations.

\section{Numerical Results}\label{sec:ph}

In the numerical analysis, we adopt LO set CTEQ6L1 \cite{Pumplin:2002vw} of proton PDFs, available through the LHAPDF6 program library \cite{Buckley:2014ana}, and the LO formula for $\alpha_s^{(n_f)}(\mu_r)$, with $n_f=4$ and $\Lambda_\mathrm{QCD}^{(4)}=215~\mathrm{MeV}$~\cite{Pumplin:2002vw}.
We set the renormalization and factorization scales to be $\mu_r=\mu_f=\xi\left(\sqrt{4M_{H_1}^2+p_T^2}+\sqrt{4M_{H_2}^2+p_T^2}\,\right)/2$ for the direct production of charmonia $H_1$ and $H_2$ with common transverse momentum $p_T$.
We take $\xi=1$ as default and vary $\xi$ between 1/2 and 2 to estimate the theoretical uncertainties from unknown higher orders in $\alpha_s$.
For definiteness, we choose $m_c=1.5~\mathrm{GeV}$ as is frequently done in similar analyses, so as to facilitate comparisons with the literature. 
The values $m_{J/\psi}=3.097~\mathrm{GeV}$, $m_{\psi(2S)}=3.686~\mathrm{GeV}$, $\mathrm{Br}(\psi(2S) \rightarrow J/\psi+X)=61.4\%$ for masses and branching fraction are taken from the latest Review of Particle Physics \cite{ParticleDataGroup:2022pth}.
As for the $J/\psi$ and $\psi(2S)$ CS LDMEs, we use the LO ones evaluated from the wave functions at the origin for the Buchm\"{u}ller-Tye potential \cite{Eichten:1995ch},
\begin{eqnarray}
  \langle \mathcal{O}^{J/\psi}({}^3S_1^{[1]})\rangle
  &=&1.16~\mathrm{GeV}^3\,,\nonumber
  \\
  \langle \mathcal{O}^{\psi(2S)}({}^3S_1^{[1]})\rangle
  &=&0.758~\mathrm{GeV}^3\,,
\end{eqnarray}
and estimate the $\mathcal{O}(v^2)$ ones by the ratio obtained for the $J/\psi$ case in Ref.~\cite{Bodwin:2006dn},
\begin{equation}\label{eq: v^2 LDMEs}
\frac{\langle \mathcal{P}^{J/\psi}({}^3S_1^{[1]})\rangle}
{\langle \mathcal{O}^{J/\psi}({}^3S_1^{[1]} \rangle}  =  
0.5~\mathrm{GeV}^2\approx  
\frac{\langle \mathcal{P}^{\psi(2S)}({}^3S_1^{[1]})\rangle}
{\langle\mathcal{O}^{\psi(2S)}({}^3S_1^{[1]})\rangle}\,.
\end{equation}
In the evaluation of feed-down contributions to differential cross sections, we
approximate the momentum of the $J/\psi$ meson from $\psi(2S)$ decay as 
\begin{equation}
p_{J/\psi} = \frac{m_{J/\psi}}{m_{\psi(2S)}}p_{\psi(2S)}\,.
\end{equation}

We are now in a position to confront our improved theoretical predictions for the cross section of prompt double $J/\psi$ hadroproduction in SPS with experimental data to investigate how the inclusion of $\mathcal{O}(v^2)$ relativistic corrections affects the agreement close to the production threshold.
Our prime observable is the $m_{\psi\psi}$ distribution in the first few bins.
Owing to the LO relation $m_{\psi\psi}=2\sqrt{4m_c^2+p_T^2}\cosh(\Delta y)$ \cite{He:2015qya}, this corresponds to small values of rapidity separation $|\Delta y|$, so that we also focus on the first few bins of the $|\Delta y|$ distribution.
Because the $m_{\psi\psi}$ distribution rapidly falls off with increasing value of $m_{\psi\psi}$ \cite{He:2015qya}, the total cross section is dominated by the threshold region.
As mentioned above, the bulk of the cross section close to threshold is due to the CS channel $gg\to 2c\bar{c}(^3S_1^{[1]})$, while the CO contributions are suppressed there \cite{He:2015qya}.
In fact the CO contributions to the total cross sections of direct and prompt double $J/\psi$ hadroproduction under LHCb experimental conditions at $\sqrt{s}=7$~TeV \cite{LHCb:2011kri} only amount to 8.2\% and 5.9\%, respectively \cite{He:2015qya}.

Of all available experimental data \cite{D0:2014vql,LHCb:2011kri,LHCb:2016wuo,CMS:2014cmt,ATLAS:2016ydt,LHCb:2023ybt,LHCb:2023wsl}, we select for our dedicated comparisons the recent LHCb data of Ref.~\cite{LHCb:2023ybt} because they are most precise and the SPS contribution is separated from the DPS one, which is not included in our predictions.
Reference~\cite{LHCb:2023ybt} is based on a data sample corresponding to an integrated luminosity of 4.2~fb${}^{-1}$, refers to prompt double $J/\psi$ hadroproduction at $\sqrt{s}=13$~TeV, with both $J/\psi$ mesons in the ranges $0<p_T<14$~GeV and $2.0<y<4.5$, and specifies total cross sections and various distributions.
Besides the $m_{\psi\psi}$ and $|\Delta y|$ distributions in their lower ranges, also the distribution in the rapidity $y$ of either $J/\psi$ meson is of interest for our purposes because it is also dominated by the small-$m_{\psi\psi}$ region.
In all these cases, LO NRQCD, at $\mathcal{O}(\alpha_s^4)$, is expected to yield useful approximations, to be improved by the inclusion of relativistic corrections of relative order $\mathcal{O}(v^2)$.
In the following, we only consider SPS data from Ref.~\cite{LHCb:2023ybt}.

We start by considering the total cross section and compare the LHCb result \cite{LHCb:2023ybt},
\begin{equation}
\sigma_{\mathrm{LHCb}}^{\mathrm{SPS}} = 7.9 \pm 1.2 \pm 1.1~\mathrm{nb}\,,
\label{eq:lhcbsps}
\end{equation}
where the first error is statistical and the second one systematic, with our LO and $\mathcal{O}(v^2)$-corrected NRQCD predictions,
\begin{eqnarray}
  \sigma^{\mathrm{LO}}_{\mathrm{NRQCD}}&=&17.43^{+2.44}_{-3.98}{}~\mathrm{nb}\,,
\nonumber\\
\sigma^{\mathrm{NLO}}_{\mathrm{NRQCD}}&=&12.02^{+1.98}_{-2.81}{}~\mathrm{nb}\,,
\end{eqnarray}
respectively.
As is familiar from similar comparisons in the literature \cite{He:2021oyy}, the LO NRQCD prediction greatly overshoots the experimental result, by more than a factor of 2.
This overshoot is significantly reduced by the inclusion of the $\mathcal{O}(v^2)$ corrections, to about 50\%.

\begin{figure}[htbp]
\begin{tabular}{ll}
  \includegraphics[width=0.49\textwidth]{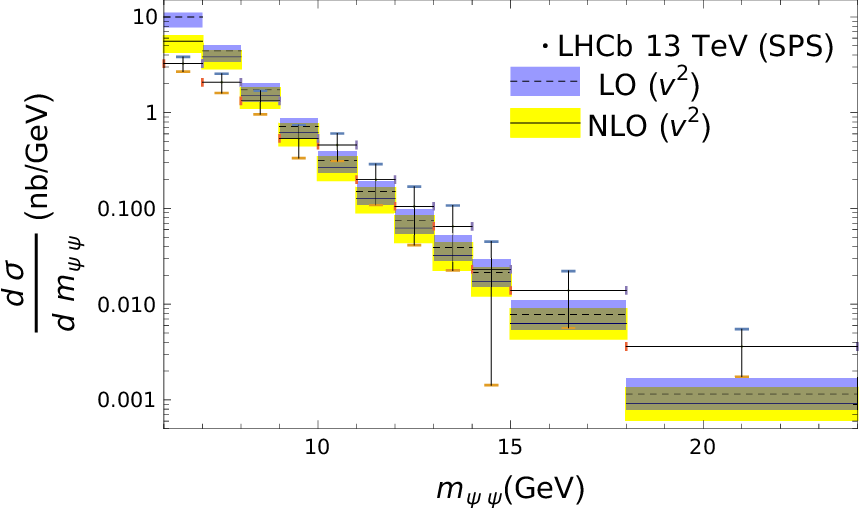} &
  \includegraphics[width=0.49\textwidth]{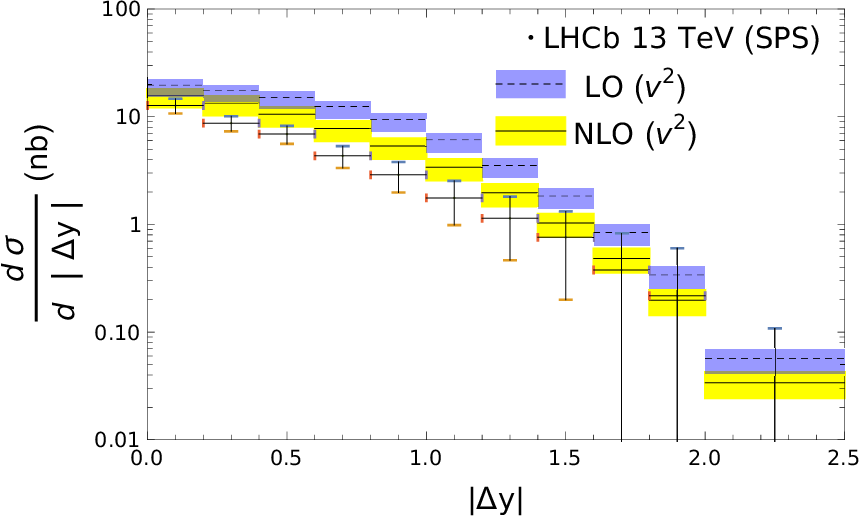}\\
  (a) & (b)\\
\includegraphics[width=0.49\textwidth]{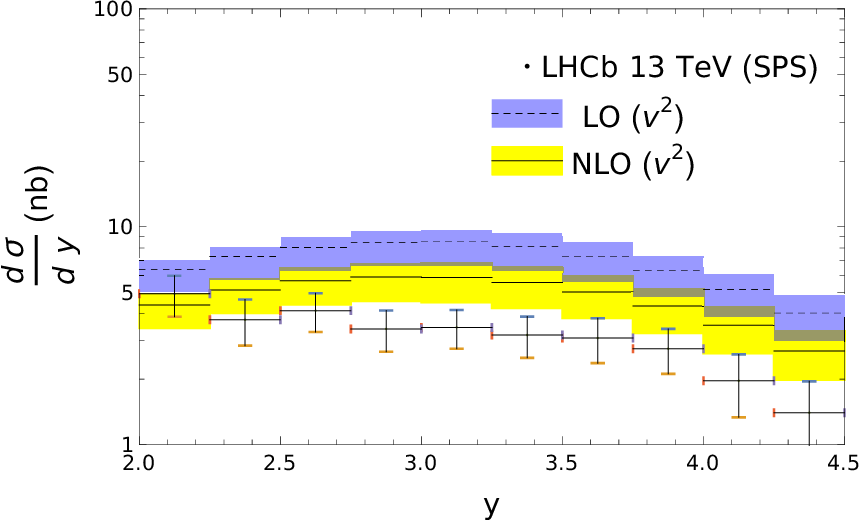} &\\
(c) &
\end{tabular}
  \caption{\label{fig:psipsi}
    The (a) $m_{\psi\psi}$, (b) $|\Delta y|$, and (c) $y$ distributions of prompt double $J/\psi$ hadroproduction in SPS measured by LHCb \cite{LHCb:2023ybt} at $\sqrt{s}=13$~TeV are compared with our CS NRQCD predictions at LO (dashed lines) and NLO with respect to relativistic corrections (solid lines).
The theoretical uncertainties are indicated by the shaded blue (LO) and yellow (NLO) bands.}
\end{figure}

We now turn to the $m_{\psi\psi}$, $|\Delta y|$, and $y$ distributions, which are
shown in Figs.~\ref{fig:psipsi}(a)--(c), respectively, where the LHCb data \cite{LHCb:2023ybt} are compared with our LO and $\mathcal{O}(v^2)$-corrected NRQCD predictions.
As explained above, theory-to-data comparisons for the $m_{\psi\psi}$ and $|\Delta y|$ distributions are meaningful only for the first few bins, where the CS contribution is dominant.
From Figs.~\ref{fig:psipsi}(a)--(c), we observe that the $\mathcal{O}(v^2)$ corrections are throughout negative.
As may be gleaned from Fig.~\ref{fig:psipsi}(a), they are most significant close to threshold, being $-44$\% in the first bin $6~\mathrm{GeV}<m_{\psi\psi}<7~\mathrm{GeV}$, and monotonically become smaller in magnitude as $m_{\psi\psi}$ increases, reaching $-20$\% in the last bin $18~\mathrm{GeV}<m_{\psi\psi}<24~\mathrm{GeV}$.
As is clear from the discussion in Sec.~\ref{sec:frame}, the relativistic corrections are composed of the SDC contribution, from $G_{1,2}({}^3S_1^{[1]},{}^3S_1^{[1]})$, and the residual contribution, from the phase space.
Detailed analysis reveals that the latter is dominant in the first $m_{\psi\psi}$ bin, while $G_{1,2}({}^3S_1^{[1]},{}^3S_1^{[1]})$ wins out in higher bins.
In the first few $m_{\psi\psi}$ bins, the inclusion of the $\mathcal{O}(v^2)$ corrections significantly improves the agreement between NRQCD predictions and LHCb data.
The same is true for the $|\Delta y|$ and $y$ distributions in Figs.~\ref{fig:psipsi}(b) and (c), respectively.
In fact, as for the first three bins of the $|\Delta y|$ distribution and most of the bins of the $y$ distribution, the errors of the $\mathcal{O}(v^2)$-corrected NRQCD prediction and the LHCb data overlap.

We note in passing that inclusion of our relativistic corrections also leads to improved descriptions of prompt double $J/\psi$ hadroproduction at $\sqrt{s}=7$~TeV \cite{LHCb:2011kri} and prompt $J/\psi+\psi(2S)$ hadroproduction at $\sqrt{s}=13$~TeV \cite{LHCb:2023wsl} as measured by the LHCb Collaboration, albeit without SPS-DPS separation.
To illustrate this, we merely consider the total cross sections.
Specifically, we have
\begin{eqnarray}
  \sigma_{\mathrm{LHCb}}^{\mathrm{SPS+DPS}} &=& 5.1 \pm 1.0 \pm 1.1~\mathrm{nb}\,,
  \nonumber\\
  \sigma^{\mathrm{LO}}_{\mathrm{NRQCD}}&=&9.60^{+2.57}_{-2.73}{}~\mathrm{nb}\,,
  \nonumber\\
  \sigma^{\mathrm{NLO}}_{\mathrm{NRQCD}}&=& 6.55^{+1.92}_{-1.89}{}~\mathrm{nb}\,,
  \label{eq:lhcb2011}
\end{eqnarray}
for Ref.~\cite{LHCb:2011kri} and
\begin{eqnarray}
  \sigma_{\mathrm{LHCb}}^{\mathrm{SPS+DPS}} &=& 4.49 \pm 0.71 \pm 0.26~\mathrm{nb}\,,
  \nonumber\\
  \sigma^{\mathrm{LO}}_{\mathrm{NRQCD}}&=&16.23^{+2.17}_{-3.67}{}~\mathrm{nb}\,,
  \nonumber\\
  \sigma^{\mathrm{NLO}}_{\mathrm{NRQCD}}&=&5.93^{+1.21}_{-1.44}{}~\mathrm{nb}\,,
\label{eq:lhcb2023}
\end{eqnarray}
for Ref.~\cite{LHCb:2023wsl}.
It is reasonable to expect that the SPS fractions of the LHCb results in Eqs.~(\ref{eq:lhcb2011}) and (\ref{eq:lhcb2023}) are in the same ball park as for the LHCb results in Ref.~\cite{LHCb:2023ybt}, where it is about 50\%, as may be inferred by comparing Eq.~(\ref{eq:lhcbsps}) with \cite{LHCb:2023ybt}
\begin{equation}
\sigma_{\mathrm{LHCb}}^{\mathrm{SPS+DPS}} = 16.36 \pm 0.28 \pm 0.88~\mathrm{nb}\,.
\end{equation}

\begin{figure}[htbp]
  \includegraphics[width=0.49\textwidth]{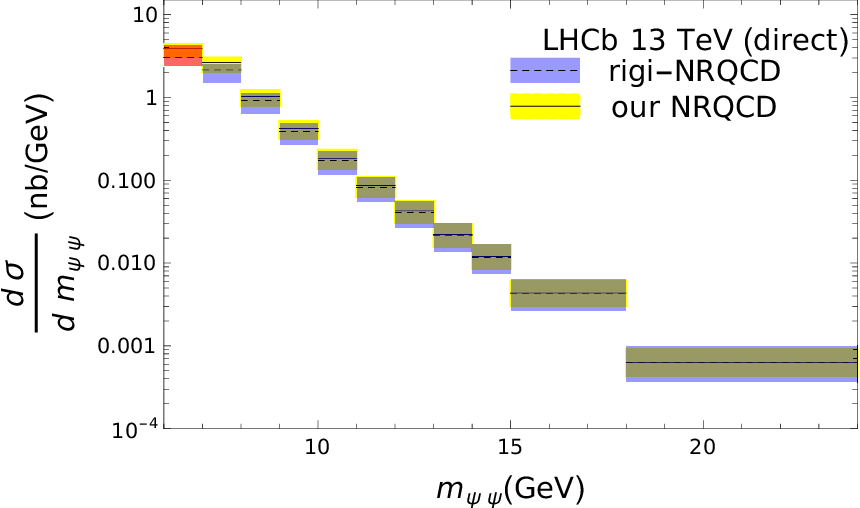}
  \caption{\label{fig:mjj_naive} 
The $m_{\psi\psi}$ distribution of direct double $J/\psi$ hadroproduction in SPS under LHCb experimental conditions evaluated in NRQCD with $\mathcal{O}(v^2)$ relativistic corrections evaluated in our modified approach (solid lines) is compared with its counterpart evaluated by rigidly expanding the phase space in $v^2$ (dashed lines).
The theoretical uncertainties are indicated by the shaded blue (rigi-NRQCD) and yellow (our NRQCD) bands.
Negative values are shown in red.}
\end{figure}

As explained in Sec.~\ref{sec:frame}, the conventional expansion of the phase space in $v^2$ diverges in the limit when the two-gluon invariant mass $\sqrt{s}$ approaches the four-quark threshold $4m_c$, which renders the naive evaluation of the relativistic corrections ill-defined for $m_{\psi\psi}$ values close to $2m_{J/\psi}$.
To overcome this problem, we proposed to evaluate the phase space using the meson mass $m_{J/\psi}$ instead of the quark mass $m_c$ and not to expand it in $v^2$.
We illustrate the numerical effect of this modification for the $m_{\psi\psi}$ distribution of direct double $J/\psi$ hadroproduction in SPS in Fig.~\ref{fig:mjj_naive}.
We observe from Fig.~\ref{fig:mjj_naive}, that the naive evaluation leads to a negative result of similar magnitude in the first $m_{\psi\psi}$ bin, violating NRQCD factorization.
This reflects the fact that, in the treatment of the phase space, the formal expansion parameter $v^2$ is endowed with the factor $8m_c^2/(s-16m_c^2)$, which is large close to threshold.
On the other hand, the relative difference between the naive and modified evaluations rapidly decreases with increasing value of $m_{\psi\psi}$ and becomes insignificant beyond 10~GeV.
A rigorous solution of this problem is likely to require a resummation to all orders of $v^2$, which is beyond the scope of this work.
We refer the interested reader to Ref.~\cite{Bodwin:2007ga}, where such a resummation was discussed for the exclusive process $e^+e^-\to J/\psi+\eta_c$.

\section{Summary}\label{sec:summary}

In summary, we studied the relativistic corrections of relative order $\mathcal{O}(v^2)$ to prompt double $J/\psi$ hadroproduction in SPS within the NRQCD factorization framework in view of previous observations \cite{Li:2013csa,He:2019qqr,He:2021oyy} that NRQCD predictions lacking such corrections significantly overshoot experimental measurements close to the $2J/\psi$ production threshold \cite{LHCb:2011kri,LHCb:2016wuo,CMS:2014cmt,ATLAS:2016ydt,LHCb:2023ybt,LHCb:2023wsl}.
We found that the conventional power counting for the expansion in $v^2$ is offset for the phase space close to threshold due to an enhancement factor of the form $8m_c^2/(s-16m_c^2)$, where $\sqrt{s}$ is the partonic center-of-mass energy.
As a cure to this problem, we proposed an alternative treatment of the phase space implemented with physical charmonium masses.

We confronted our improved theoretical predictions with recent high-precision measurements by the LHCb Collaboration, in which the SPS contributions were extracted by appropriate acceptance cuts \cite{LHCb:2023ybt}.
In fact, the inclusion of $\mathcal{O}(v^2)$ corrections turned out to significantly improve the NRQCD description of the measured $m_{\psi\psi}$ distribution in the first few bins.
As expected, this improvement was found to carry over to the $|\Delta y|$ distribution in the first few bins, to the $y$ distribution, and to the total cross section, which all receive dominant contributions from the near-threshold region of the $m_{\psi\psi}$ distribution.

In conclusion, relativistic corrections play an important role in our understanding of the mechanism underlying double charmonium hadroproduction at the LHC, especially in kinematic configurations close to the production threshold, and enhance the predictive power of NRQCD factorization.
Further improvements of the theoretical description near threshold are expected from the resummations of relativistic corrections \cite{Bodwin:2007ga} and logarithmic corrections due to multiple soft-gluon emission \cite{Ma:2017xno}, which lie beyond of the scope of this work.  

\begin{acknowledgments}
  
We thank Guang-Zhi Xu for helpful communications enabling a comparison with the results of Ref.~\cite{Li:2013csa}.
This work was supported in part by the German Research Foundation DFG through Research Unit FOR~2926 ``Next Generation Perturbative QCD for Hadron Structure: 
Preparing for the Electron-Ion Collider" under Grant No.~409651613. The work of X.B.J. was supported in part by National Natural Science Foundation of China under Grant No.~12061131006.

\end{acknowledgments}

\end{document}